\documentclass[%
superscriptaddress, amsmath,amssymb, aps, prb, twocolumn ]{revtex4-2}

\usepackage[utf8]{inputenc}
\usepackage{graphicx}
\usepackage{bm}
\usepackage{xcolor}
\usepackage[caption=false]{subfig}
\usepackage{xfrac}
\usepackage{times}

\definecolor{darkblue}{HTML}{004D6B}
\definecolor{darkred}{HTML}{8c1515}
\usepackage{hyperref}
\hypersetup{ colorlinks=true, urlcolor=darkblue, citecolor=darkred,
    linkcolor=darkblue, breaklinks }

\newcommand{\op}[1]{\hat{#1}}
\newcommand{\secref}[1]{Sec.~\ref{#1}} 
\newcommand{\Eqref}[1]{Eq.~\eqref{#1}} 
\newcommand{\figref}[1]{Fig.~\ref{#1}}
\newcommand{\appref}[1]{Appendix \ref{#1}} 

\newcommand{\vect}[1]{\textbf{#1}}
\graphicspath{{figures/}}

\begin{document}
\title{Classical Chaos in Quantum Computers}

\author{Simon-Dominik Börner}
\affiliation{Institute for Theoretical Physics, University of Cologne, 50937 Cologne, Germany}

\author{Christoph Berke}
\affiliation{Institute for Theoretical Physics, University of Cologne, 50937 Cologne, Germany}

\author{David P. DiVincenzo}
\affiliation{Institute for Quantum Information, RWTH Aachen University, 52056 Aachen, Germany}
\affiliation{J\"ulich-Aachen Research Alliance (JARA), Fundamentals of Future Information Technologies, 52425 J\"ulich, Germany}
\affiliation{Peter Gr\"unberg Institute, Theoretical Nanoelectronics,
Forschungszentrum J\"ulich, 52425 J\"ulich, Germany}%

\author{Simon Trebst}
\affiliation{Institute for Theoretical Physics, University of Cologne, 50937 Cologne, Germany}

\author{Alexander Altland}
\affiliation{Institute for Theoretical Physics, University of Cologne, 50937 Cologne, Germany}

\begin{abstract}
The development of quantum computing hardware is facing the challenge that current-day quantum processors,
comprising  50-100 qubits, already operate outside the range of quantum simulation on classical computers. In this paper we demonstrate that the simulation of \textit{classical} limits can be a potent diagnostic tool potentially mitigating this problem. As a testbed for our approach we consider the transmon qubit processor, a computing platform in which the coupling of large numbers of nonlinear quantum oscillators may trigger destabilizing chaotic resonances. We find that classical and
    quantum simulations lead to similar stability metrics (classical Lyapunov
    exponents vs.~quantum wave function participation ratios) in systems with
    $\mathcal{O}(10)$ transmons.  However, the big advantage
    of classical simulation is that it can be pushed to large systems
    comprising  up to thousands of qubits. We exhibit the utility of this classical
    toolbox by simulating all current IBM transmon chips, including the recently
    announced 433-qubit processor of the Osprey generation, as well as future
    devices with $1{,}121$ qubits (Condor generation). For realistic system parameters, we find a systematic increase of Lyapunov exponents with system size, suggesting that larger layouts require added efforts in information protection. 
\end{abstract}

\maketitle

\section{Introduction}

Coupled mathematical pendula are textbook paradigms of deterministic classical
chaos \cite{tabor1989chaos}. When excited to energies large enough that the nonlinearity of the pendulum
potential becomes sizeable, a transition from integrable harmonic motion to
chaotic dynamics generically takes place. In the world of quantum physics, the
mathematical pendulum finds a prominent realization as the transmon
superconducting qubit \cite{koch2007}, with the `gravitational potential' defined by a Josephson
junction, and the `kinetic energy' by a micro capacitor. The cosine nonlinearity of the former
is required to gap the lowest two quantum states of the transmon (aka
the qubit) against the  noncomputational higher lying parts of the spectrum in a
nonresonant manner \cite{krantzQuantumEngineerGuide2019,blais_circuit_2021}. Coupled transmons/pendula define the brickwork of superconducting
quantum processors \cite{kjaergaard_superconducting_2020}.  On the basis of quantum-to-classical correspondence, one may
suspect traces of chaotic dynamics -- which are toxic where quantum
computing is concerned -- to be visible in this setting \cite{orell_probing_2019,berke_transmon_2022}. Indeed, they are, and
there appear to be two master strategies for keeping them out: decouple qubits
off-operation by so-called tunable couplers \cite{yan_tunable_2018,xu_high-fidelity_2020,chen_qubit_2014} (an approach applied in, e.g.,  Google's Sycamore quantum chip \cite{aruteQuantumSupremacyUsing2019a}), or intentionally detune
the oscillator frequencies of neighboring qubits relative to each other, to
avoid  dangerous resonances (as done in current quantum chips by the IBM \cite{hertzberg_laser-annealing_2021} / Delft \cite{krinnerRealizingRepeatedQuantum2022} / ETH Z\"urich  \cite{versluis_scalable_2017} consortia).

Both approaches have their individual advantages. The first reliably stabilizes
the system, but at the expense of substantial overhead  hardware for switchable
coupling \cite{aruteQuantumSupremacyUsing2019a}. The second avoids this complication, but instead  introduces
engineered \emph{disorder}. (In the parlance of quantum many-body physics, the
ensuing state of matter is called `many-body localized' \cite{huse_phenomenology_2014, abanin_colloquium_2019}. In it, the system
becomes effectively integrable, but at the expense of site to site randomness,
with perhaps unintended side effects in large-scale structures.)

In this paper, we investigate manifestations of \emph{classical} chaos in transmon
arrays, tuned to a classical limit by setting $\hbar=0$ \cite{classicalchaosblais}. Otherwise, our systems
-- their transmon frequencies, coupling strength, system layout, etc. -- are
modeled in agreement with published data for existing quantum chips \cite{IBMQuantumCloud,zhangHighfidelitySuperconductingQuantum2020}. Why
would one enforce a classical limit upon a quantum computer? Our prime
motivation for this study is that the transmon array displays a highly developed
quantum-to-classical correspondence: exact diagonalization performed for the
corresponding  \emph{quantum} systems show quantum chaos in  parametric
regions with classical chaos, and its absence in regions without. For systems
with up to ten transmons (the limit for our quantum calculations), this
correspondence is developed with high accuracy. The point now is that the
analysis of the classical limit can be pushed  to  $\mathcal{O}(10^3)$
resonators, i.e.,\ numbers comparable to those of state-of-art processors
deployed in cloud computing services \cite{IBMQuantumCloud}, and way beyond anything that can be
quantum simulated on a classical computer. Our study of \emph{large-scale}, but \emph{static} transmon storage devices complements the existing literature on the link between classical chaos and the driven quantum dynamics in circuit QED set-ups \cite{classicalchaosblais,pietikainenA,pietikainenB}. While these earlier works study the effects of nonlinearities in circuits subject to additional complexity, e.g., with drive lines for the implementation of gates, they focus on \emph{small-scale} architectures (single qubit coupled to a cavity).

Our construction of classical dynamics simulations as a diagnostic toolbox for
large scale processor layouts is organized in three steps. After a quick review
of current day transmon hardware in Section~\ref{sec:Hardware}, we present an
analysis of classical chaos in linear arrays of  two to ten transmons in
Section~\ref{sec:Classical}. A principal observation is that chaos is present already in the two-transmon context but only at excitation energies way beyond those relevant for quantum applications. For ten transmons, however, manifestations of chaos bleed down into the excitation range corresponding to that of the quantum computational qubit Hilbert space. We take this observation
as an incentive for a thorough comparison of classical and quantum dynamics for ten transmon arrays in Section~\ref{sec:Predictive}. The number `10' is special
 inasmuch as it defines the maximal number of transmons for which we can run
precision quantum simulation with good statistics 
\footnote{The quantum simulations must include states outside the computational subspace that are fully intermingled with the qubit states \cite{Boerner2023, berke_transmon_2022}. The simulations are therefore numerically costly, see also the discussion in \appref{app:algscaling}.}.

Focusing on classical Lyapunov
exponents and many-body wave function statistics as prime indicators of
classical and quantum chaos, respectively, we will construct a comparison chart
showing the predictive potential of classical simulation. Specifically, we will
argue that Lyapunov exponents measuring the instability of the classical system
are, in a statistical sense, in quantitative correspondence to the quantum
system's inverse participation ratios (IPR). The latter are a measure for the spread
of quantum wave function over Fock space, and provide microscopic information on
the integrity of qubits \cite{EversAndersonTransition}. 

In Section \ref{sec:state-of-the-art} we then turn to the trump card of the
classical approach, the option to simulate arrays of up to thousands of
transmons,  including realistic transmon wiring \cite{chamberlandTopologicalSubsystemCodes2020} and other hardware design
elements. Specifically, we will simulate transmon chips contained in the current
IBM roadmap \cite{ExpandingIBMQuantum2021}, from the 27-transmon Falcon  to the announced 1,021-transmon Condor chip.
Assuming that the quantum-to-classical correspondence observed at 10-transmon
level extends to larger qubit numbers, this analysis yields valuable insights
into the design of (future) processor layouts. We will consider advanced design principles, as realized in frequency-engineered cross resonance architectures \cite{rigetti_fully_2010},
where  IPRs close to unity -- representing perfect single transmon wave function localization -- can be reached by engineered fine tuning \cite{zhangHighfidelitySuperconductingQuantum2020, hertzberg_laser-annealing_2021}. Our classical
analysis will demonstrate the manner in which  the Lyapunov exponents signal the
proximity to such sweet spots. At the same time, they show a systematic tendency
to increase for larger system architectures, which we take as indication that
maintaining the stability of these sophisticated designs will require an additional
engineering effort. We conclude in section \ref{sec:Summary}.

\section{Transmon hardware}
\label{sec:Hardware}

Transmon-based quantum computers are among the most developed information
processing platforms of the NISQ era \cite{Preskill2018NISQ} and have been used
in several recent experimental landmarks: the first demonstration of quantum
computational advantage \cite{aruteQuantumSupremacyUsing2019a}, the simulation
of topologically ordered states
\cite{satzingerRealizingTopologicallyOrdered2021}, and small instances of
error-correcting experiments with surface code logical qubits
\cite{krinnerRealizingRepeatedQuantum2022,acharyaSuppressingQuantumErrors2023}. While there are other promising
approaches based on superconducting circuits at the level of single qubits or
few-qubit devices (e.g., the fluxonium
\cite{baoFluxoniumAlternativeQubit2022,fluxonium1} or the C-shunt flux qubit
\cite{yanFluxQubitRevisited2016}), transmons are the clear front-runner when it
comes to integrating $\mathcal{O}(50)$--$\mathcal{O}(100)$ qubits into a single
viable processor. This property makes the transmon the preferred choice for
applications where scalability is paramount, e.g., the recent demonstration of
the performance improvement of a logical qubit with the surface code distance,
conducted in a 72-qubit device \cite{acharyaSuppressingQuantumErrors2023}.
Processors containing more than $1{,}000$ transmon qubits are expected to be
launched in upcoming years. For example, IBM's quantum roadmap announces a
monolithic processor with $1{,}121$ qubits for the year 2023 and a modular
quantum computer with $4{,}158$ qubits in 2025 \cite{ExpandingIBMQuantum2021}.

In the following, we review the transmon qubit array  and introduce a model which stays close to the systems used in reality. We then proceed to address the main question of this paper: what can we learn from the \emph{classical} physics of this system  about the functioning of the \emph{quantum} processor?

\subsection{Transmons}

In its simplest form, a single transmon consists of
only a single Josephson junction and a large shunting capacitance. Its
Hamiltonian is given by \cite{koch2007}
\begin{align}
		\op{H}_\text{Tr} = 4 E_C  \op{n}^2 - E_J \cos \op{\varphi} \,,
        \label{eq:singletransmon}
\end{align}
where $\op{n}$ is the charge operator~\footnote{One may generalize the charge operator $\op{n}\to \op{n}-n_g$ to include an offset $n_g$
describing the influence of an external gate
voltage or of environmental charge fluctuation. However, for our purposes, these effects are of little relevance.} counting the number of Cooper pairs that
have traversed the junction and $\op{\varphi}$ is the superconducting phase
conjugate to $\op{n}$, i.e., $\left[ \op{\varphi}, \op{n} \right] = i$. The
Josephson energy $E_J$ is a macroscopic parameter describing the ability of
Cooper pairs to pass the tunnel barrier, and $E_C$ is the charging energy
necessary to transfer one electron through the junction.  $E_C$ is
proportional to the total capacitance of the circuit and -- due to the sizeable
shunting capacitance  -- can be made small compared to $E_J$, which pushes the
dimensionless parameter $E_J/E_C$ to the transmon regime where  $E_J/E_C \gtrsim
20$. Typical values of $E_C$ range from 100~MHz to 400~MHz
\cite{blais_circuit_2021}, while $E_J$ often lies near 12.5~GHz (note here that we give energies in Hz, by setting $h$ to 1).
The ground
state and the first excited state serve as the two qubit states $|0\rangle$ and
$|1\rangle$. The energy spacing between the two qubit states $h\nu_{01}$
typically takes values of $\nu_{01} \equiv \nu_q = 5$~GHz, where  $\nu_q$ is called the
qubit frequency.  

\begin{figure}[t]
    \centering
    \includegraphics{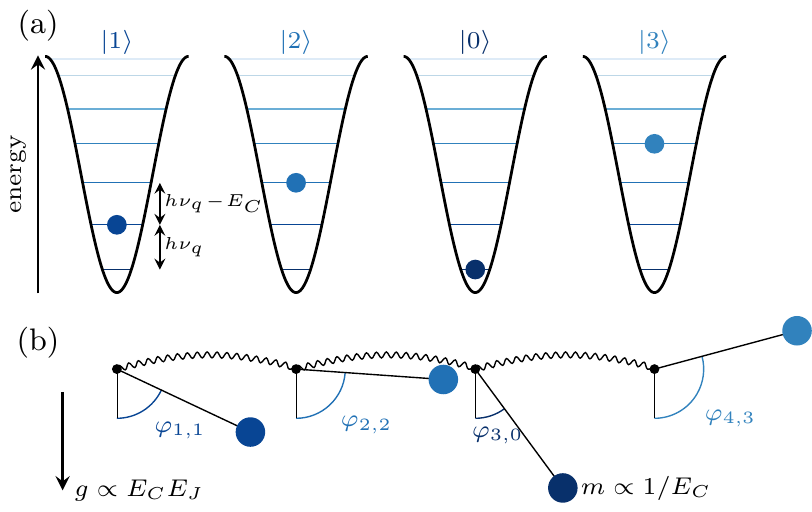}
    \caption{\textbf{Quantum vs. classical transmon array.} 
    (a) A chain of four transmons initialized in the quantum state  $|1203\rangle$ of the corresponding $\cos$-potentials, where the integers, $i=1,2,0,3$ correspond to the bound state energies $E_i$.  (b) The corresponding classical rotors, initialized at angular deflections corresponding to the  energies $E_i$  as discussed in the main text,  and  
    the springs
    connecting the suspension points representing the  angular
    momentum coupling. 
     }
    \label{fig:pendulums}
\end{figure}

For our
purposes, it will be advantageous to consider $\op{\varphi}$ as an angular variable
with conjugate angular momentum  $\op{L}_z = \hbar \op{n}$. With the identification $E_C = \hbar^2 /8ml^2$, and $g = E_J/ml = 8E_CE_Jl/\hbar^2$,  Eq.~\eqref{eq:singletransmon} then describes a quantum pendulum (see Fig.~\ref{fig:pendulums})
\begin{equation}
 \op{H}_\text{Tr}=\frac{\op{L}_z^2}{2ml^2} - mgl \cos \op{\varphi} 
\end{equation}
 of mass $m$, rigid length $l$, and gravitational
constant $g$.

The \emph{coupling} of neighboring transmons is often realized via a capacitive interaction  $\op{n}_i\op{n}_j$ between
their charge degrees of freedom. The full Hamiltonian of an array of 
coupled transmons then reads \cite{gambettajuelich}
\begin{align}
	\op{H} = 4 E_C \sum_i \op{n}_i^2 - \sum_i E_{J,i} \cos \op{\varphi}_i + T \sum \limits_{\langle i,j \rangle} \op{n}_i \op{n}_j \,. \label{eq:fullmodel}
\end{align}
Here, the site dependence of the Josephson energies, $E_{J,i}$, accounts for unavoidable 
fabrication imprecisions, usually of the order of 5\% to 10\%
\cite{rosenblattVariabilityMetricsJosephson2017,gambettaBuildingLogicalQubits2017}. 
While these tolerances can be reduced by post-processing or by advanced fabrication techniques (see Ref.~\cite{berke_transmon_2022} for a detailed discussion)  frequency variations 
are often  introduced intentionally to detune neighboring transmons  during
`gate-off' times and in this way suppress undesired correlations
\cite{barends_superconducting_2014}. (Adjustable Josephson energy variations
are usually realized via  so-called flux tunable
transmons where a single Josephson interface is replaced by a SQUID~\cite{barends_superconducting_2014}.)

By comparison, variations of both the charging energy $E_C$ (here assumed to be at the
value 250~MHz) and of the coupling energies $T$ ($\sim 30$ MHz for flux tunable
transmons\cite{barends_superconducting_2014} and $\sim
3$--$5$ MHz \cite{sheldon_procedure_2016} for single-junction `fixed-frequency')
are of lesser relevance and will be ignored throughout. We also will not
consider the important concept of tunable couplers \cite{yan_tunable_2018},
i.e., additional hardware allowing to vary the  coupling on operation at the expense of extra noise sources.

\subsection{Classical limit}

Transmon quantum computing relies on the deep quantization of the Josephson junctions'  two lowest energy levels. However, as we are going to demonstrate in the following, the system's classical limit -- a network of classical pendula -- contains valuable information about the physics of the transmon array.
In this limit, the operators $\op{\varphi}_i, \op{n}_i$ are demoted to real valued variables
$\varphi_i,n_i$, and the commutation relation $\left[ \op{\varphi}_i, \op{n}_j
\right] = i \delta_{i,j}$ turns into a Poisson bracket
\begin{equation}
	\lbrace  \varphi_i, n_j \rbrace = \delta_{i,j} \,.
\end{equation}
Hamilton's canonical equations of motion then read
\begin{align}
	  \dot{\varphi}_i &= \lbrace  \varphi_i, H \rbrace = \frac{\partial H}{\partial n_i} = 8 E_C n_i + T \sum \limits_{j=\text{NN}(i)} n_j \label{eq:dphi}\,, \\
	 \dot{n}_i &= \lbrace n_i, H \rbrace  = - \frac{\partial H}{\partial  \varphi_i} = - {E_{J,i}} \sin \varphi_i \label{eq:dn} \,,
\end{align}
where $H$ is the classical Hamilton function obtained by replacing
$\op{\varphi}$ and $\op{n}$ with their classical counterparts in the Hamiltonian
$\op{H}$ in \Eqref{eq:fullmodel}, and the sum is over  nearest neighbors of
transmons. These equations describe a system of classical  pendula with a
`momentum-momentum' interaction arising from the capacitive coupling.


To mimic a transmon  initialized in one of its eigenstates $|0\rangle$,
$|1\rangle$, $|2\rangle$, \ldots\, we first compute the energies $E_a$ of the
quantum model, where $a=0,1$ for computational states. We then
initialize the classical rotor in a phase space configuration
$(n,\varphi)=(0,\varphi_a)$, where  $-E_J \cos \varphi_a=E_a$, or
\begin{equation}
	\varphi_a = \arccos \left(- \frac{E_a}{E_J} \right). \label{eq:phiinit}
\end{equation}
In other words, the classical pendulum is started in a configuration of maximal
potential and zero kinetic energy, see Fig.~\ref{fig:pendulums}. As illustrated,
the quantum transmon supports 7 bound states, for the chosen parameters of  $E_J=12.5$~GHz and $E_C=250$~MHz,
 and we distinguish between as many classical
initial configurations. 

\section{Classical chaos}
\label{sec:Classical}

Coupled nonlinear pendula are a paradigm of deterministic chaos, and the question to be addressed in this paper is to what extent the corresponding instabilities also affect the quantum array. To approach this question, we first consider simple toy models:  two coupled pendula, and the generalization to a chain of $L$ of them. 

\subsection{Two coupled transmons}

Chaotic behavior already emerges in the classical two-transmon Hamiltonian,
provided that the system is excited to sufficiently high
energies~\cite{borner2020}. In this reduced setting, the phase space spanned by
the coordinates $(n_1,n_2,\varphi_1,\varphi_2)$ is four dimensional, implying
that the onset of chaos can be demonstrated via the powerful concept of Poincaré
sections, i.e., stroboscopic images defined by the crossing of classical
trajectories on the three-dimensional surface of conserved energy with
the two-dimensional surface defined by the fixation of one of the coordinates.
To be specific, we here keep track of the pairs $\left( \varphi_1, n_1 \right)$
at $\varphi_2 = 0$ and $n_2 > 0$, where the second condition fixes a sense of
traversal.

\begin{figure}
    \centering
    \includegraphics[width=\columnwidth]{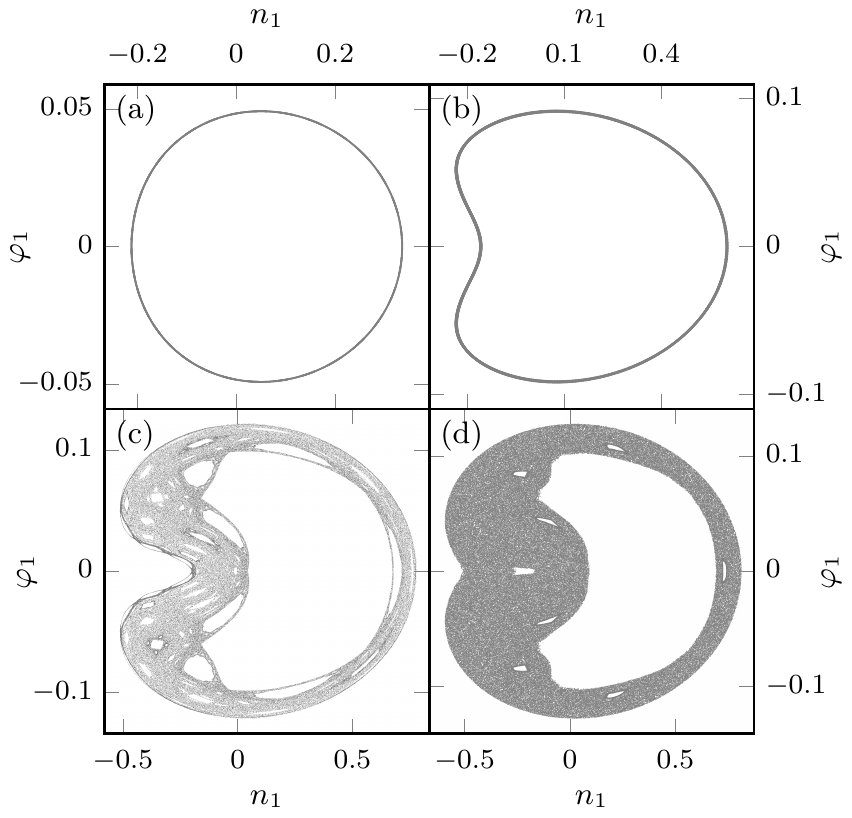}
    \caption{\textbf{Poincar\'e sections for a system of two coupled transmons.}
    Shown are Poincar\'e sections in the $\varphi_1$--$n_1$ plane with
    $\varphi_2 = 0$ and $n_2>0$. The transmons are initialized with
    $\varphi_1(t=0)=n_2(0)=0$, $n_1(0)=0.01$ and $\varphi_2 = \pi - x$, where
    (a) $x=0.1$ (b) $x =0.05$, (c) $x = 0.02$ and (d) $x = 0.0005$. 
    We set $T=40$ MHz and $E_C = 300$ MHz.
     The Josephson energies ($E_{J,1}=98.8$ GHz and $E_{J,2} = 101.2$ GHz) 
     lie above the experimentally relevant parameter range. }
    \label{fig:twotransmons}
\end{figure}

\begin{figure}
    \centering
    \includegraphics[width=.99\columnwidth]{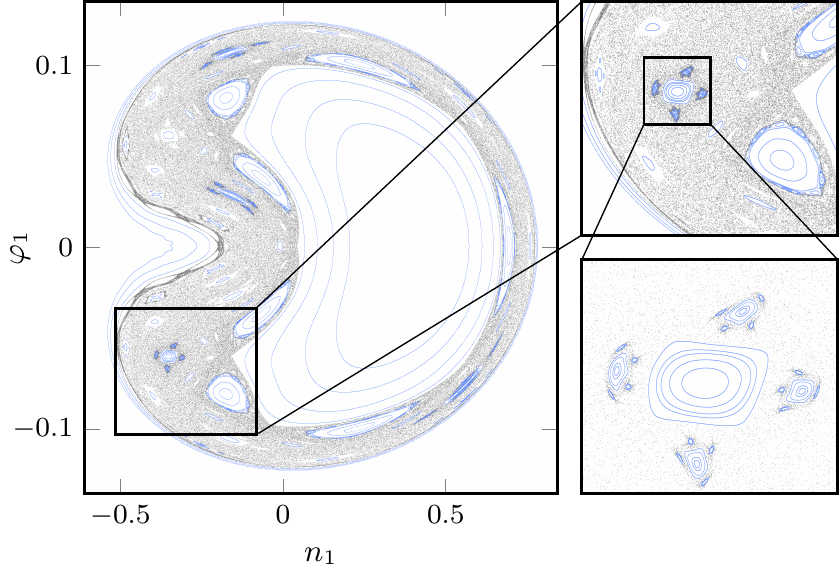}
    \caption{\textbf{Signatures of nonlinear dynamics in a system of two coupled
    transmons}. Parameters are chosen as in Fig. \ref{fig:twotransmons}(c). The
    figures represent  orbits with identical energy but different initial
    conditions, leading to integrable (blue) or chaotic (gray) dynamics. Note
    the self similar structure of orbits and satellite orbits upon
    magnification.}
    \label{fig:poincare2}
\end{figure}

To monitor the onset of irregular dynamics, we vary the initial angle
$\varphi_2^\text{init}$ while keeping $\varphi_1^\text{init} = n_2^\text{init} =
0$ and $n_1^\text{init} = 0.01$ fixed. \figref{fig:twotransmons} shows four
different Poincar\'e sections for initial angles $\varphi_2^\text{init}$,
which -- from (a) to (d) -- get progressively closer to $\pi$. While the closed
curves for (a) and (b) indicate that the motion is (quasi)periodic and thus
integrable, one observes a qualitative change upon further increasing
$\varphi_2^\text{init}$.  The Poincar\'e section then extends over a finite
fraction of the $\varphi_1$-$n_1$ plane, as is expected for nonintegrable
systems. We have also confirmed that in the non-integrable regions of the Poincaré plot there is exponential sensitivity to initial conditions, as witnessed by finite Lyapunov exponents. 
\figref{fig:poincare2}, which is a fine-grained representation of the section (c)
in \figref{fig:twotransmons}, indeed shows various textbook signatures \cite{Timm2022} of a
system whose phase space contains integrable and chaotic regions. Examples of these include 
Kolmogorov-Arnold-Moser (KAM) tori \cite{poschelLectureClassicalKAM2009}, the
intermittent presence of elliptic and hyperbolic fixed points required by the
Poincar\'e-Birkhoff theorem (see, e.g., Ref.~\cite{Ketzmerick2021}), and
self-similarity. 

While the above analysis is proof of principle of the presence of chaos in the two-transmon system, it is of no practical relevance: The energies where the onset of chaos is observed lie well beyond those relevant for computing applications, i.e.,\ the energies corresponding to the computational states $|00\rangle,|10\rangle,|11\rangle$ according to the mapping discussed in the previous section. 
However, as we are going to show next, the situation changes dramatically when we pass from two- to many-transmon arrays. 

\begin{figure}[b]
    \centering
    \includegraphics[scale = 1]{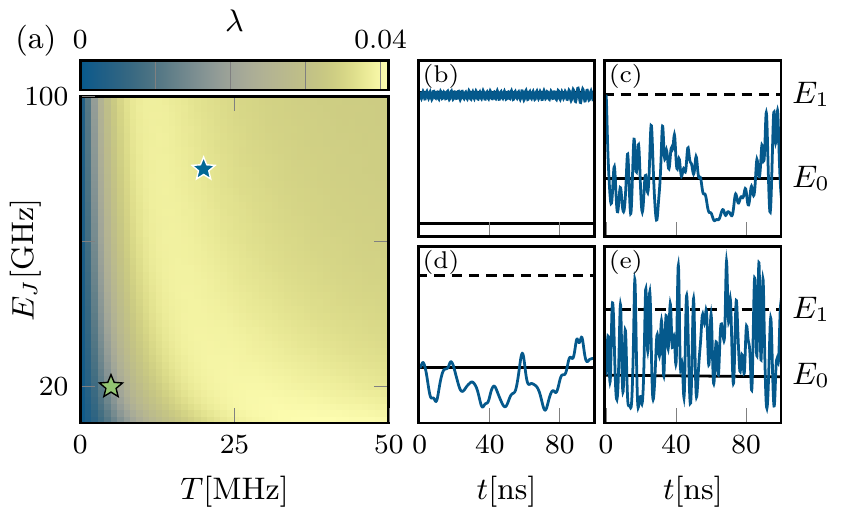}
    \caption{\textbf{Classical chaos in a chain of ten coupled transmons.} (a)
	The maximal Lyapunov exponent in the $(E_J,T)$  plane averaged over at least $8{,}000$ disorder realizations. (b)--(e) Time-dependent single transmon energies for the sites 5 (upper row) and 6 (lower row) for two disorder
	realizations. The left (right) column corresponds to the parameters marked
	by the green (blue) star in (a). Whereas
	the Hamilton functions remain near their initial values in (b) and (d), they
	fluctuate heavily on timescales much shorter than typical decoherence times
	in (c) and (e). Only in the first case can one draw 
 a credible conclusion that the initial bitstring is
 `$1010\dots$' from the energies at $t>0$. This
	consideration shows that the magnitude of $\lambda$, which is
	small for (b) and (d) but large for (c) and (e), can serve as a quality
	indicator of the classical transmon `storage device'.}
    \label{fig:tentransmons}
\end{figure}

\subsection{Ten coupled transmons} \label{subsec:tentransmons}

As a first step towards understanding  the physics of many-transmon arrays, we
now discuss  a model of $L=10$ transmons coupled in a chain geometry, for
energies pertinent to quantum computing applications. More precisely, the system
is prepared in the $|1010\dots\rangle$ state, i.e., the angles $\varphi_i$ on
the even (odd) sites are chosen such that the initial single transmon energies
correspond to the quantum mechanical energies $E_0$ ($E_1$). To diagnose chaos,
we calculate the maximal Lyapunov exponent $\lambda$, i.e.,\ the rate at which
trajectories with initial phase space distance $\delta \pi$  diverge, i.e.,
$\delta \pi (t) \approx \delta \pi \exp (\lambda t )$,  for more details, see
\appref{app:methods}.

\figref{fig:tentransmons}(a) shows the results as a function of the Josephson
energy $E_J$ and the coupling $T$, averaged over a large number of `disorder'
configurations. Each of these instances is generated by the independent drawing of ten values $E_{J,i}$ from a normal distribution with mean $E_J$ and standard deviation  $\delta E_J = \sqrt{E_J E_C/8}$. As detailed in
\appref{app:disorder}, this peculiar choice for $\delta E_J$ ensures a constant
\emph{frequency} disorder (as $E_J$ varies) of $\delta \nu_q \approx
\tfrac{E_C}{2}$ akin to what is found for current-generation quantum
processors.

\begin{figure}
    \centering
    \includegraphics[scale = 1]{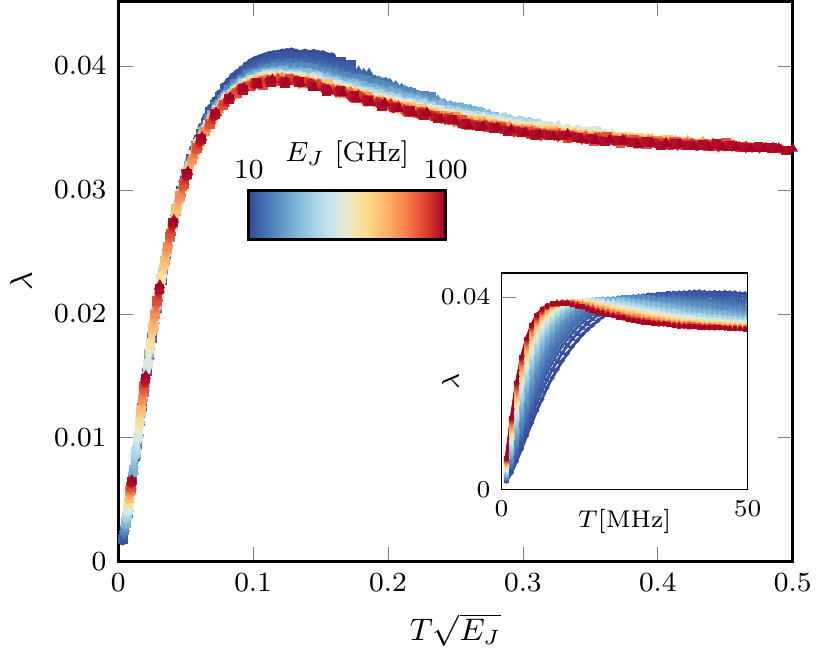}
    \caption{\textbf{Data collapse for the maximal Lyapunov exponent.} 
		Lyapunov exponent plotted for different values of $E_J$ as a function of $T$ (inset) and of the scaling variable $T\sqrt{E_J}$ main panel. 		In the latter case the data collapses almost perfectly for small $T$ and still reasonably well for larger values: $T\sqrt{E_J}$ is the relevant 	parameter controlling the onset of chaos. }    
 	\label{fig:classical_scaling}
\end{figure}

The omnipresence of chaos for
 experimentally relevant parameter values reveals itself in
 a non-vanishing Lyapunov exponent for almost the entire phase diagram of
  \figref{fig:tentransmons}(a). The exception to the rule is a narrow
 region near $T=0$,  the limit of uncoupled pendula. Increasing $T$ leads to a
 sharp increase of the Lyapunov exponent towards a maximum value, and finally
 the levelling at a value slightly below that maximum.
 \figref{fig:classical_scaling} shows this behavior of the $T$-dependent
 Lyapunov exponent, now plotted as a function of the scaling variable
 $T\sqrt{E_J}$ for different values of $E_J$. The oberservation here is that
 under this rescaling, the two parameter function $\lambda(T,E_J)\to
 \lambda(T\sqrt{E_J})$ shows data collapse, indicating that $T\sqrt{E_J}$ is the
 relevant parameter controlling the onset of (quantum) chaos. We will return to
 this point when we discuss the quantum interpretation of our classical findings
 in the next section.

\begin{figure}
    \centering
    \includegraphics[scale = 0.95]{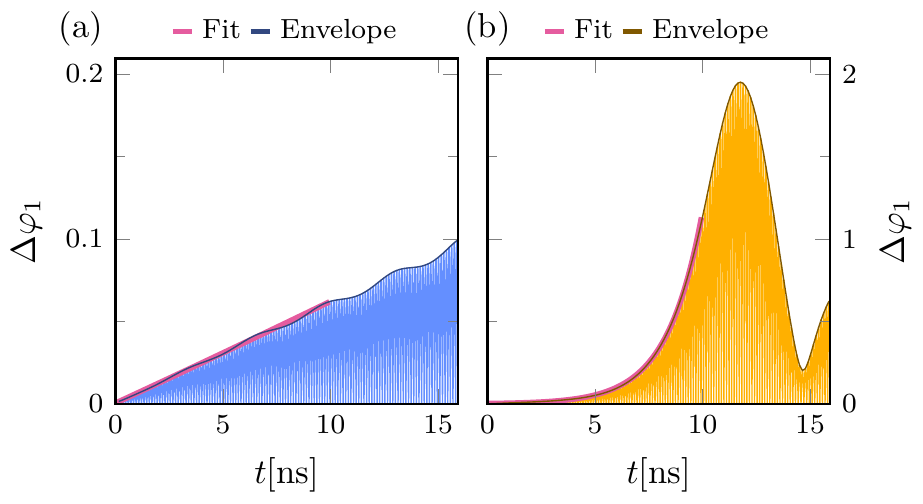}
    \caption{\textbf{Divergence of classical trajectories.} Shown is the difference $\Delta \varphi_1$ between the angular coordinates of
    the first qubit in a 10-qubit chain for two trajectories initialized with a
    starting mismatch $\varphi_1'(0)=1.001\varphi_1(0)$. The two trajectories are
    initialized to mimic  a $|1010101010\rangle$ quantum state via their angular  
    displacement. System parameters: (a) $T=5$ MHz, $\delta E_J=0.5$ GHz, (b)
    $T=10$ MHz, $\delta E_J=0.1$ GHz, and in both cases $E_C=250$ MHz and $E_J\approx 10$ GHz.
    The magenta lines show exponential fits to the envelopes of $\Delta \varphi_1$.
    For (a) the fit stretches out to a linear function, while for (b) the fit gives a strong exponential growth of $\Delta\varphi_1$.
    The corresponding maximal Lyapunov exponents are (a) $\lambda\approx0$ and (b) $\lambda\approx0.03$.}
    \label{fig:ExpFig}
\end{figure}

To develop some intuition for the meaning of a non-vanishing Lyapunov exponent, Fig.~\ref{fig:ExpFig} exemplifies the sensitivity to variations in initial conditions (a mismatch of $0.1\%$ in the first angular coordinate of a 10 transmon array) for realistic system parameters. While for $\lambda=0$ (left) the initial mismatch increases linearly, we observe exponential behavior on a scale magnified by one order of magnitude for $\lambda \approx 0.03$, which eventually gives way to aperiodic fluctuations due to the compact range of the angular parameter space.

What are the implications of these findings for the application of the transmon
array as an information processing device? Specifically, the reliable storage of
information requires that a qubit initialized in either of the computational
states $|0\rangle$ or $|1\rangle$ maintains this state under the evolution
governed by the time independent Hamiltonian Eq.~\eqref{eq:fullmodel}. In the
classical reading, this situation corresponds to a transmon initialized in one
of the energies $E_{0,1}$ matching the qubit energies. The maintenance of the
state translates to the condition that the time dependent energy
$E_i(t)=H_i(t)$, i.e.,\ the instantaneous value of the $i$th transmon's Hamilton
function, remain close to its initial value. (We note that the total energy of
the array is dynamically conserved, but that of its constituent transmons is
not.) At the very least, it should not cross $E_0$ if initialized in $E_1$ and
vice versa. 

Panels \figref{fig:tentransmons} (b), (c), and (d), (e)  show the energies $E_5$
and $E_6$, respectively,   for an array initialized in a configuration with
energies $(E_1,E_0,E_1, \dots   )$ corresponding to the quantum state
$|1,0,1,\dots \rangle$. The left and right panels correspond to parameter values
marked by a green and blue star in panel (a). We observe that for near-integrable 
dynamics (green), the initial energies $E_5(0)=E_1$ and $E_6(0)=E_0$
remain approximately conserved. In the chaotic case (blue), however, there are
erratic fluctuations, exceeding the energy spacing $E_1-E_0$. These fluctuations
build up after a few nanoseconds, far shorter than characteristic qubit
coherence times. We conjecture, and will discuss in more detail below, that in
this regime the functioning of the storage is compromised.

\begin{figure}
    \centering
    \includegraphics{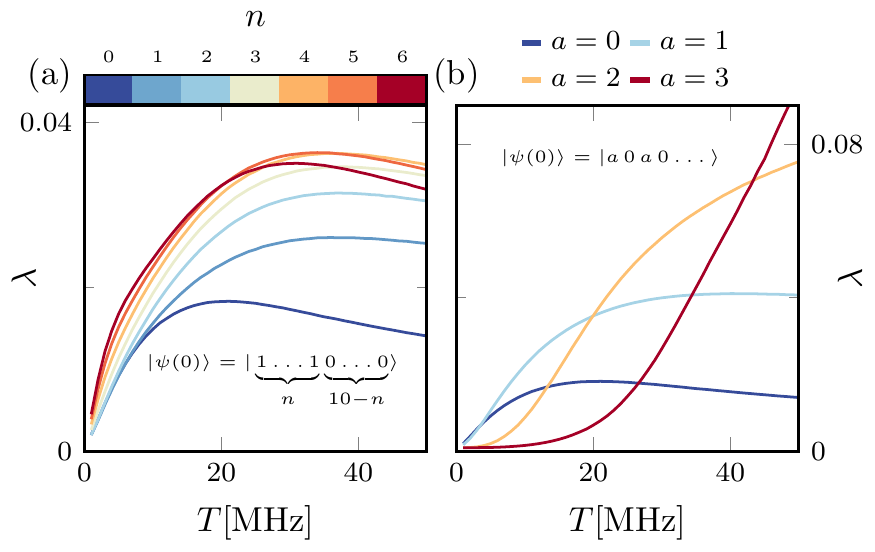} 
    \caption{\textbf{Influence of the state energy on the dynamics.} (a) 
    Lyapunov exponents for computational states of increasing energy.
     (b) Lyapunov exponents for configurations including transmons initialized in states $E_{0,1,2,3}$ including values outside the computational sector.  Values given are from a $10$ qubit transmon chain
    with $E_C=250$ MHz, $E_J=10$ GHz, and $\delta E_J=559$ MHz. Each point is
    the mean value of $20{,}000$ disorder realizations. For discussion, see text.
    }
    \label{fig:energydensity1}
\end{figure}

\figref{fig:energydensity1} (a) shows the generalization of the Lyapunov data to
different transmon configurations. In the main panel, we plot
$\lambda$ for a variety of (classical analogs of) computational
states. The data shows a general trend towards larger Lyapunov exponents for
increasing state energy, i.e., larger numbers of $E_1 \leftrightarrow |1\rangle$
initializations. All curves exhibit the same qualitative behavior as a function
of $T$ as that discussed above, the reaching of a maximum value followed by
saturation. Panel (b) shows data for states, $(E_a,E_0,E_a, \dots)$, with
$a=0,1,2,3$. (The generalization to non-computational states, $a>1$, is
practically relevant as transmon gate operations transiently couple to states
outside the computational sector \cite{Boerner2023}.) Two features stand out: The Lyapunov
exponents reach (i) larger values, however, these maximal values are attained (ii) only
for larger values of the coupling. We reason that the relatively higher inertia to changes in $T$ has to do with the fact that for larger energies of individual transmons the coupling represents a relatively weaker perturbation. We note that these findings are consistent with a recent study \cite{mansikkamakiHardCoreBosonsTransmon2022} of a (quantum) Bose-Hubbard model, that finds a suppression of the effective interaction between states with large occupation numbers on individual transmon sites.


\section{Predictive power of classical simulations}
\label{sec:Predictive}

At this point, we have  discussed key signatures of the classical
dynamics of small scale transmon arrays. The big question now of course  is what
bearings these findings have for our actual subject of interest, the quantum
processor. In this section, we formulate an answer in a succession of steps.
First, as a warmup, we show that several of the observations of the previous
section afford a quantum interpretation. We then compare  our results above
with those of quantum simulations for the  ten transmon array, and for identical
material parameters (except that now $\hbar\not=0$, of course), to observe a
high level of agreement: classical chaos implies quantum chaos, and vice versa.
We finally turn to the  trump card of the classical approach, namely the option
to reliably simulate arrays of thousands of transmons. Assuming that the
quantum-to-classical fidelity extends to large numbers, we thus have a tool to
obtain stability measures for realistic quantum hardware outside the reach of
quantum simulation on classical computers. In section
\ref{sec:state-of-the-art}, we substantiate this point by simulating large scale
two-dimensional transmon arrays of current IBM design. 

\subsection{Quantum to classical correspondence (qualitative)}

Quantum mechanically, the passage from integrable to chaotic dynamics upon
increasing $T$ is a manifestation of a Fock space (de)localization transition:  
In the transmon regime, $E_J \gg E_C$, \Eqref{eq:fullmodel} is well approximated
by the attractive Bose-Hubbard model \footnote{To obtain this representation,
one passes from number and phase to an oscillator basis, $(\hat n,\hat{\varphi})\to \sqrt{\hat n}\exp(i \hat \varphi)\equiv a$. An expansion of the
$\cos$-potential near its minimum up to quartic order then yields the Hubbard
Hamiltonian $\hat{H}(a,a^\dagger)$ .}. Thinking of the Fock basis, defined by
the occupation numbers of the transmons $\left( n_1, n_2,\ldots,n_L \right)$, as
a lattice whose  sites are connected through the capacitive interaction, one
expects that wave functions delocalize if the hopping amplitude $t$ between
these lattice sites is larger than the `on-site' (in Fock space) energy
difference $\Delta\epsilon$. 
In terms of the transmon array parameters, the hopping amplitude reads $t = T\sqrt{E_J}/\sqrt{32E_C}$ \cite{berke_transmon_2022,
blais_circuit_2021}. 
The many-body level spacing depends on the total anharmonicity of the Fock space lattice sites and the disorder in the qubit frequencies. In our
simulations, both contributions are proportional to $E_C$ and independent of
$E_J$, see \appref{app:disorder}, i.e., $\Delta \epsilon \propto E_C$.  
For the scaling variable, this yields $\tfrac{t}{\Delta \epsilon} \propto T\sqrt{E_J}/\sqrt{E_C^3}$. Since $E_C$ is kept constant in the simulations, one expects that the contour lines separating regimes of (integrable) many-body localized and extended chaotic regimes, scale as $\tfrac{t}{\Delta \epsilon} \propto T\sqrt{E_J} =
\mathrm{const}$. Below, we will demonstrate this scaling for the system's  wave
function statistics. The finding that the classical Lyapunov exponents scaled
with the same parameter is consistent with the paradigm that quantum and
classical chaos condition each other. 

\begin{figure}
    \centering
    \includegraphics[width=0.5 \textwidth]{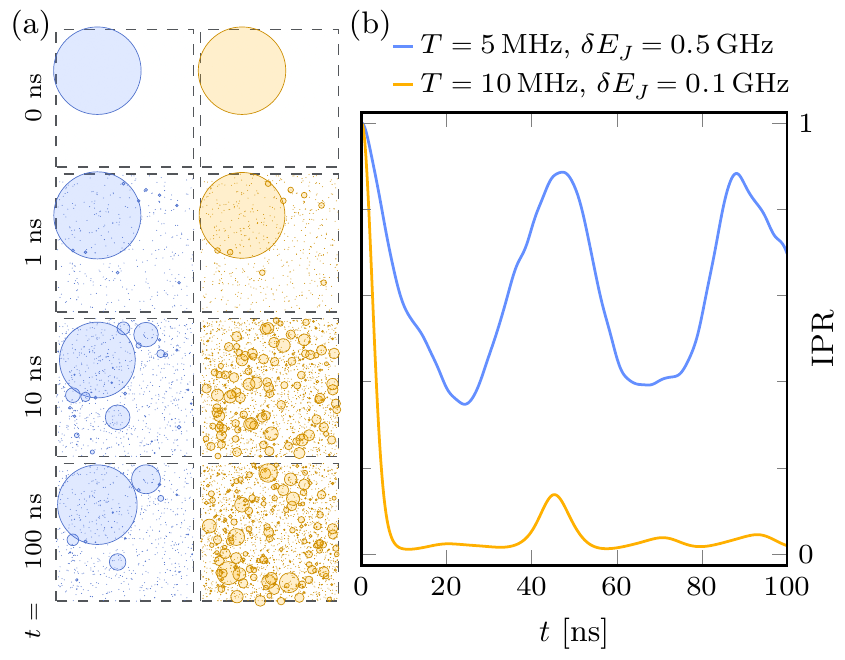}
    \caption{\textbf{Visualization of Fock space state fragmentation} in the time
    evolution, $|\psi(t)\rangle$, of a state initialized as $|\psi(0)\rangle =
    |1010\ldots\rangle$, i.e. the quantum state corresponding to the classical
    initial condition considered in Fig.~\ref{fig:ExpFig}. The system paramters,
    $E_J,E_C,T$, too, are those previously used in Fig.~\ref{fig:ExpFig} (a),
    blue, integrable and (b), yellow, chaotic. (a) Circles are centered around
    randomly chosen assignments of occupation number states to coordinates in
    the plane, and their areas quantify the probability to find
    $|\psi(t)\rangle$ in these states. The four rows illustrate how the
    integrable (chaotic) state retains its structure (fragments) in a succession
    of four discrete time steps. (b) The  continuous time evolution of the
    inverse participation ratio of $|\psi(t)\rangle$. }
    \label{fig:delocalizingwavefunction}
\end{figure}

In the classical context, `delocalization' is delocalization away from the
integrable orbits of the oscillator motion of individual transmons. Quantum
mechanically, it stands for the spreading of many body wave functions over a large
set of occupation number sites. To illustrate this phenomenon, we consider the
quantum evolution of the states corresponding to the classical initial
conditions discussed in connection with Fig.~\ref{fig:ExpFig}. The resulting
Fock space structure is visualized in Fig.~\ref{fig:delocalizingwavefunction}
(a), where the the circles are centered around an arbitrary mapping of Fock
space sites of total occupation number $5$ (e.g. $|01130 \dots\rangle$) to
the two-dimensional plane, and circle areas quantify the square amplitude of
the states at these sites. The left (right) panels map four stages in the time
evolution of an initial state with parameters previously used in the left
(right) panel of the classical Fig.~\ref{fig:ExpFig}. We observe that a
vanishing (large) Lyapunov exponent corresponds to approximate state
stationarity (fragmentation) in the quantum system. In the following, we discuss
inverse participation ratios as a means to quantify these structures.

\subsection{Quantum to classical correspondence (quantitative)}

In the following, we consider the wave function inverse participation ratio (IPR) as a sensitive measure of quantum chaotic dynamics \cite{PhysRevLett.112.057203}. 
For a many-body  wave function $|\psi \rangle$, this quantity is
defined as 
\begin{align} 
    \mathrm{IPR} = \sum\limits_k |\langle k | \psi \rangle|^4 \,,
\end{align}
where the sum is over the Fock state basis. The limiting cases to be distinguished are IPR $\approx 1$ indicating localization in the $k$-basis, and  IPR = $1/\mathrm{dim}\, \mathcal{H}$ for chaotic states ergodically  
spread over Hilbert space \cite{EversAndersonTransition}. 

\begin{figure}
    \centering 
    \includegraphics[width=\columnwidth]{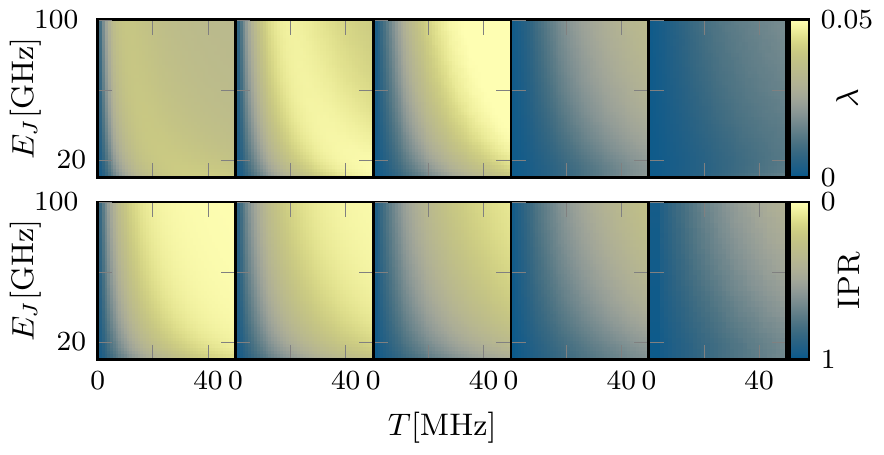}
    \caption{{\bf Quantum to classical correspondence.}
    Comparison of the classical and the quantum dynamics of ten coupled
    transmon oscillators, averaged over at least $3{,}000$  realizations of disorder of increasing  strength $\delta
    \nu_q = c \cdot E_C$
    with (left to right), c = 1/2 (as in \figref{fig:tentransmons}), c = 1, c = 2, c = 4, c = 6. The desired frequency disorder is realized by scaling $\delta E_J \propto \sqrt{E_J}$, as discussed in \appref{app:disorder}.
    For the classical simulation, the  system is initialized in the
    $(E_1,E_0,E_1,\ldots)$, in the quantum case, the IPR is
    averaged over states with total Fock space occupation number $L/2$.}
    \label{fig:2Dvardis}
\end{figure}

The bottom left panel of \figref{fig:2Dvardis} color-codes the IPR for the ten
transmon quantum array in the occupation number eigenbasis of the $T=0$ system,
where dark blue  and  bright yellow encode the above limiting cases of localization
and ergodicity, respectively.  
The upper left panel shows the previously computed Lyapunov exponents in the
same representation. The two measures evidently  show  similar behavior as a function of the material
parameters. In particular, the lines of constant IPR/$\lambda$  
both follow the parametric $E_J \propto 1/\sqrt{T}$ dependence, as discussed above.

The remaining panels extend this comparison to larger values of the disorder,
from the `natural disorder' in the left column, as also discussed in \figref{fig:tentransmons}, to about ten times larger
disorder, $\delta \nu_q > 1$~GHz, realized, e.g., in recent flux-tunable architectures
\cite{krinnerRealizingRepeatedQuantum2022}. The main point here is that, at
first counterintuitively, disorder may support integrable dynamics: Upon increasing disorder the chaotic regions retract and eventually vanish. The physics behind this observation is that
increasing disorder means a diminished susceptibility for  the transmons to be driven
into a chaotically resonant regime by transmon coupling. 

We finally remark that the presence of a shallow maximum of the classical Lyapunov exponent at intermediate $T$
observed in the last section is consistent with the
proposal \cite{selsDynamicalObstructionLocalization2021a} of a domain of
\emph{maximal chaos} in-between the localized and the ergodic regime. The
statement is that  in  transit from  integrable to chaotic phases one passes a
regime with exponentially enhanced eigenvector susceptibility. Eigenstates in
this intermediate  terrain, although not yet fully extended over Hilbert space, show higher sensitivity to perturbations than in the
`usual' ergodic quantum chaotic phase. This behavior may be the quantum manifestation of an intermediate regime of exceptional classical Lyapunov sensitivity.

\begin{figure}
    \centering
    \includegraphics[scale=1]{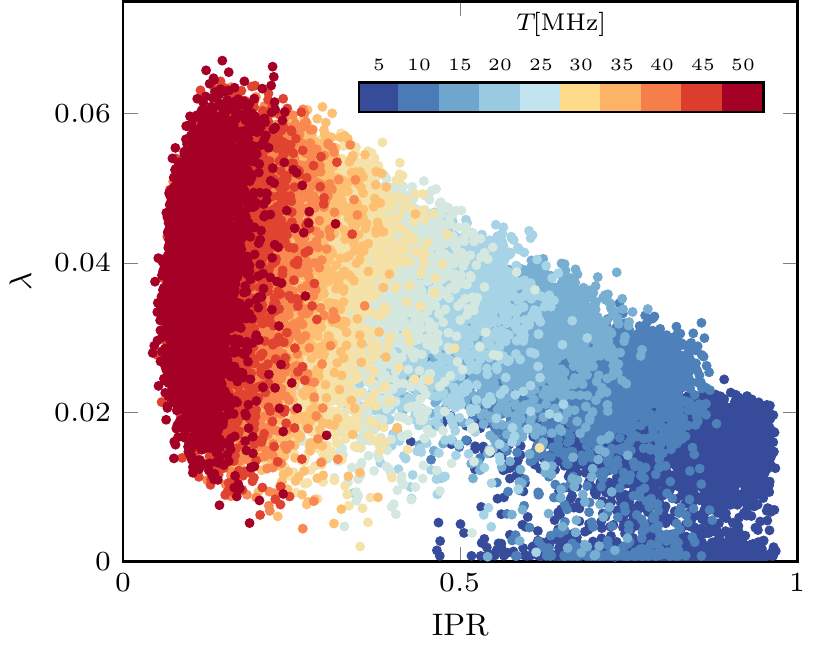}
    \caption{
    {\bf Correlation plot of maximal Lyapunov exponent and IPR.} Each of the dot
    represents one individual disorder realization for a 10 qubit transmon chain
    for different coupling strengths (indicated by the colors). The figure
    contains $2{,}000$ data points per coupling strength. We set $E_J=12.5$~GHz
    and $E_C=250$~MHz. The IPRs are averaged over all relevant wave functions,
    as explained in \figref{fig:2Dvardis}.}
    \label{fig:LyapnonovIPR-Correlation-Single-Disorder-Realisations}
\end{figure}

The above discussion qualitatively demonstrates quantum-to-classical
correspondence in the parameter space $(E_J,T,\delta E_J)$. However, ultimately,
one would like to  turn the  classical analysis into a prognostic tool for,
e.g.,  optimal quantum system parameters. To this end, the relationship between
Lyapunov exponent and IPR -- yardsticks  for classical and quantum chaos,
respectively -- needs to be understood in quantitative terms. As a first attempt
in this direction,
\figref{fig:LyapnonovIPR-Correlation-Single-Disorder-Realisations} displays the
correlations between the Lyapunov exponent and the IPR for 
coupling strengths stepwise increased between 5 and 50 MHz (color code). For each
value, we explicitly show $2{,}000$ distinct disorder realizations (each corresponding to a single point in the figure).

The  take-home message of this analysis is that the relationship between the
two quantifiers of chaos is statistical in nature. For example, a single shot
numerical measurement of a small exponent $\lambda_\mathrm{max}=0.02$ can be
consistent with IPRs distributed almost over the full range, and hence is of
not much predictive value. However, extreme value statistics applied to a large
set of values of $\lambda$ obtained for different disorder
realizations does produce valuable information. The well developed linear bound
visible in the figure implies a quantitative relation between the largest
Lyapunov exponent of the ensemble and the expected maximal value of the IPR.

\begin{figure}[t]
    \centering 
    \includegraphics[scale=0.98]{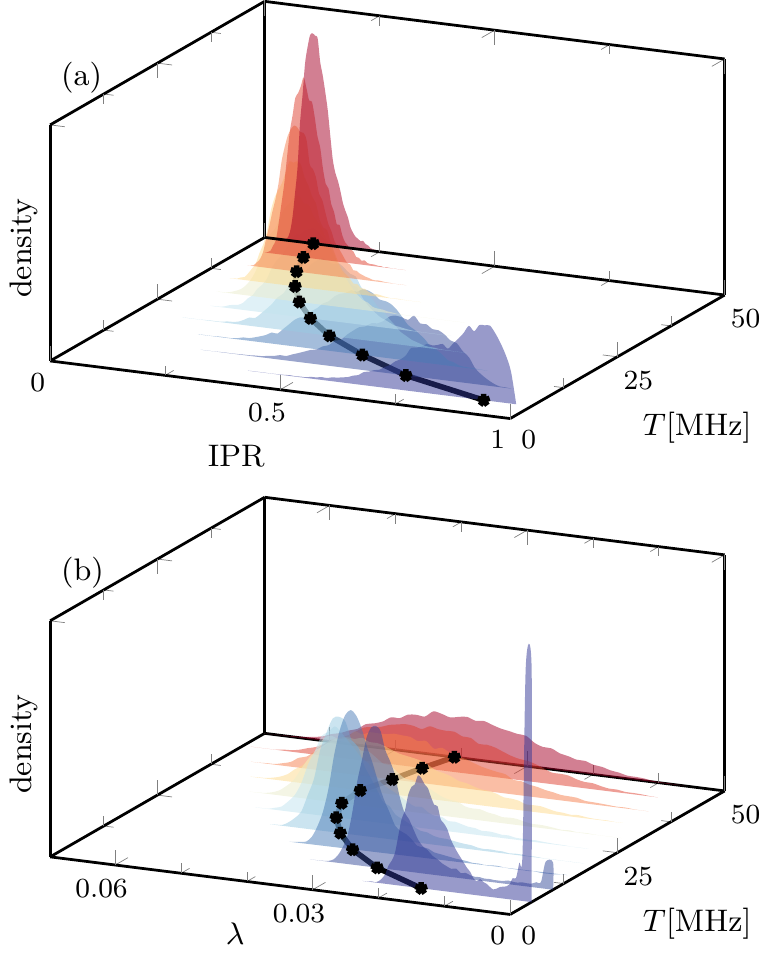}
    \caption{\textbf{Histograms of inverse participation ratio (a) and
    maximal Lyapunov exponent (b)} of Fig.
    \ref{fig:LyapnonovIPR-Correlation-Single-Disorder-Realisations}, for
    different coupling strengths. The black lines indicate the position of the
     peak maxima.} 
    \label{fig:LyapunovIPR-Correlation}
\end{figure}

We further note that, with the exception of the smallest values of the coupling,
the  IPRs show lesser statistical variation than the Lyapunov exponents. This
feature shows in the vertical stripe-like pattern visible in
\figref{fig:LyapnonovIPR-Correlation-Single-Disorder-Realisations}, and in an
alternative representation in \figref{fig:LyapunovIPR-Correlation}. That figure
shows the distribution of the measured values of Lyapunov exponents (lower
panel) and IPRs (upper panel). We observe that for large values of the coupling,
the IPRs are comparatively narrowly distributed. For smaller values, the
distribution widens, but even there remains benign in the sense that average
value and width of the distriubtion are of the same order.

We conclude  that knowledge of a distribution of Lyapunov exponents contains
information on the  spread of quantum wave functions over the transmon Hilbert
space. It is
probably safe to say that  IPRs larger than $1 /2$ are required to safeguard
the integrity of quantum storages. (Current IBM efforts (see Section \ref{sec:state-of-the-art}) strive to
reach values close to the optimal value  unity.) Our analysis shows that, for all array realizations considered in this paper, this conservative
estimate translates to the condition $\lambda<0.04$.
In practical terms, the need to harvest Lyapunov ensembles to establish these upper bounds for the IPR is not a
big issue. As discussed in the next section the computation of exponents
including for systems  way beyond current NISQ era extensions is relatively
effortless.

\subsection{Simulation of large arrays}

\begin{figure}[t]
    \centering
    \includegraphics[scale = 1]{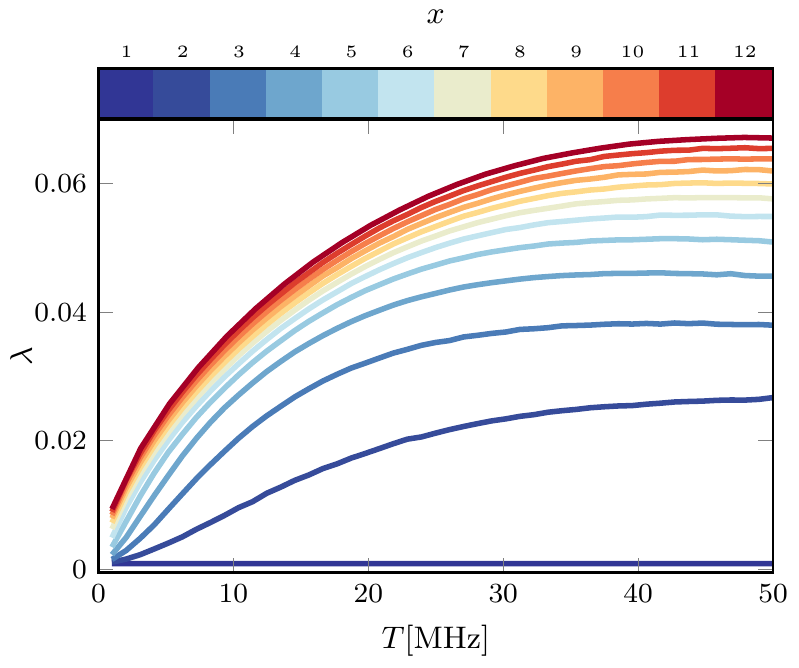}
 	\caption{\textbf{Classical chaos in chain geometries of varying length.}
	Lyapunov exponents for linear arrays of $L = 2^x,~x=1,\ldots,12$ transmons.
	($E_J=10$ GHz and disorder strength, $\delta \nu_q=\tfrac{E_C}{2}$,  the arrays are  initialized in
    the  $|E_1,E_0,E_1,\ldots\rangle$ state.) The number of
	disorder realizations varies from $2{,}000$ ($x=12$) to $20{,}000$
	($x=1$--$6$).} 
	\label{fig:classical_largersystems}
\end{figure}

The discussion so far was formulated for an array of ten transmons, a system
size comfortably in reach of both classical and quantum simulation (on classical computers).
The computational cost of quantum simulations grows exponentially in system size, limiting it to  system sizes of perhaps twice or thrice that
value, but not much larger. Currently exisiting transmon hardware with 50-100
qubits can no longer be simulated on classical machines. Our discussion above
underpins that this may be an actual limitation.  Tendencies to instability and
chaos increase with system size (for more on this, see below), and conclusions
drawn on the quantum simulation of a sub-unit of a transmon array may not fully
capture the physics of the whole.

The computational cost of classical simulation, on the other hand, grows only linearly in size, implying that arrays up to and beyond
current hardware designs are comfortably within reach. For more details on the algorithmic scaling and actual compute times of our
classical simulations we refer to \appref{app:algscaling}. 
\figref{fig:classical_largersystems} shows the
disorder-averaged Lyapunov exponent $\lambda$ for chain geometries
between 2 and $2^{12}=4{,}096$  transmons.  We observe a tendency to
more pronounced symptoms of classical chaos at larger system size. Beginning with the integrable (flat line) two-transmon arrays, the Lyapunov exponents show increasingly   sharp increase at larger $L$. 
In the next section,
we will apply the classical analysis to transmon architectures modeled after
existing hardware layouts, including two-dimensional geometries.

\section{State-of-the-art transmon chips}
\label{sec:state-of-the-art}

We now move away from the linear transmon chains studied so far to two-dimensional geometries and, in particular, a
selection of recently introduced processor generations. Our conclusions will include suggestions for
future design modifications. 


\subsection{Large-scale IBM transmon chips}
\begin{figure}[t]
    \centering
    \includegraphics{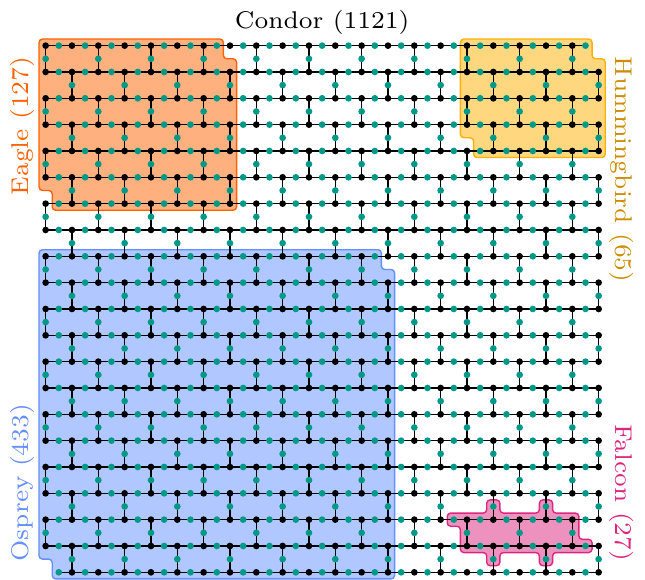}
    \caption{\textbf{Heavy-hexagon geometry and IBM processors}.  
    A $1{,}121$-transmon heavy hexagon layout with control (green) and target
    (black) qubits.  The colored cut-outs indicate the growth of  IBM's
    monolithic processor families, starting with the 27-qubits Falcon (pink,
    introduced in 2019), continued by Hummingbird (yellow, 65 qubits, 2020),
    Eagle (orange, 127 qubits, 2021), and Osprey (blue, 433 qubits, 2022), and Condor (entire lattice,
    $1{,}121$ qubits, announced for 2023).}
    \label{fig:ibmchipsgeometry}
\end{figure}

We study classical transmon dynamics on the \emph{heavy-hexagon lattice}, a
design  proposed by IBM as advantageous when upscaling the number of qubits in
cross-resonance architectures
\cite{chamberlandTopologicalSubsystemCodes2020,hertzberg_laser-annealing_2021}.
Its layout consists of a hexagonal qubit lattice with an additional transmon on
each edge, as shown in \figref{fig:ibmchipsgeometry}. (The cross-resonance
two-qubit gate involves a target and a control qubit whose correlation is
microwave activated. The above geometry with target qubits at nodes connected
via control qubits on the links of the lattice is tailored to this design
principle.) The colored segments illustrate the evolution of IBM's processor
families according to their  quantum roadmap
\cite{ExpandingIBMQuantum2021}, from Falcon (27 qubits, introduced in 2019) to  the $1{,}121$-qubit
Condor chip (to be introduced in 2023).

The main difference compared with our previous analysis is that now we are considering a two-dimensional geometry with a higher transmon connectivity (which in our simulations, however, does not imply a serious setback in  computational reach). We use the same Gaussian disorder distribution as before, and choose initial configurations of intermediate energy density within the computational subspace. To be specific, we will monitor the fate of two initial states, one with all control qubits initialized in $E_1$ and targets in $E_0$, the other one with exchanged roles $E_0 \leftrightarrow E_1$. Since there are more control than target qubits (asymptotically by a factor $3 / 2$), we refer to the former (latter)  as the high-$E$  (low-$E$) configuration.       

\begin{figure}[b]
    \centering
    \includegraphics[width=\columnwidth]{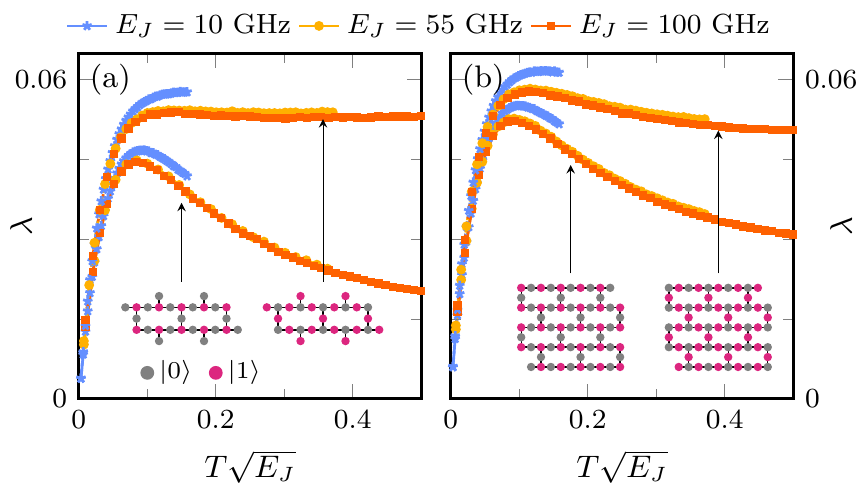}
    \caption{\textbf{Lyapunov exponents of Falcon and Hummingbird processors.} 
    Shown is a disorder-averaged Lyapunov exponent $\lambda$ for (a) the Falcon and (b) the Hummingbird processor geometry, respectively. Data is for three different values of $E_J$, and states initialized in the high (solid) and low (dotted) energy configuration. ($E_C=250$~MHz.)}
    \label{fig:ibmdevices_initcon}
\end{figure}

In \figref{fig:ibmdevices_initcon}, we compare the  Lyapunov exponent for
these two distinct density patterns for (a) the Falcon and (b) the Hummingbird
layout and three different values of the Josephson energy $E_J$. The  data shows
many commonalities with that of the chain geometry: single parameter scaling in the variable $T\sqrt{E_J}$, a steep increase towards a maximum, followed by a gradual diminishing towards a plateau value, and larger chaotic instability for states of higher energy.  
The Lyapunov exponents of the two processors are similar. Both are larger by
about $30\%$ than those of the linear geometry, i.e.,\ the extension to two
dimensions leads to a noticeable, but not dramatic increase in chaoticity. The earlier reaching of  threshold values  such as $\lambda \simeq 0.04$ implies that for larger chips one may need to work with smaller inter-transmon coupling to guarantee stability.

\begin{figure}
    \centering
    \includegraphics{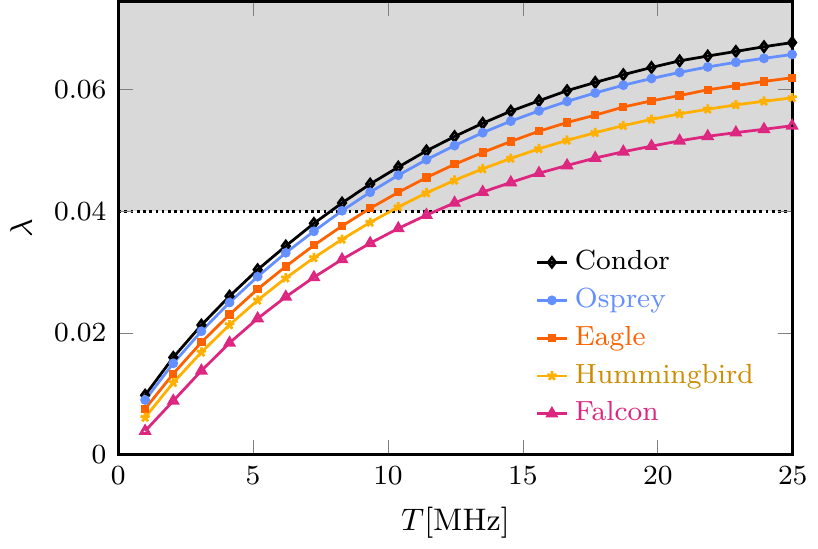}
    \caption{
        \textbf{Classical chaos in IBM quantum processors.} 
        Lyapunov exponents of states initialized in the high-$E$ configuration  
        averaged over $\sim 700$ (Condor) to $\sim 15{,}000$ (Falcon) realizations for all layouts 
        shown in  \figref{fig:ibmchipsgeometry}. ($E_J=10$~GHz and $E_C=250$~MHz.)
        }
    \label{fig:ibmchips_lyap}
\end{figure}

In \figref{fig:ibmchips_lyap}, we extend the analysis to all monolithic IBM
processors and show $\lambda$ as a function of the coupling $T$ for
each of the geometries shown in \figref{fig:ibmchipsgeometry}, and for the arrays initialized in the high-$E$ state. (The data for low-$E$ initialization looks qualitatively similar.)
 The tendency towards dynamic  instabilities rises with the total number of qubits 
 in the chip, e.g.\, the threshold value of $\lambda = 0.04$ (indicated by the dotted line)
 is reached for smaller inter-transmon coupling $T$ with increasing chip size, 
 decreasing by about a factor of two between the Falcon and Condor chips.
 This calls for additional engineering efforts to avoid chaotic instabilities 
 in larger transmon chips.

\subsection{Frequency-engineered transmon arrays}

The fidelity of cross-resonance processor layouts may be increased by
introducing engineered frequency patterns via the recently introduced
laser-annealing technique (LASIQ) \cite{hertzberg_laser-annealing_2021}. The
idea behind such patterning is similar to that of engineered disorder. A
detuning of neighboring qubit frequencies avoids degeneracies -- so-called
frequency collisions -- and their unwanted resonance effects. However, there
needs to remain a residual random frequency spread (see Ref.~\onlinecite{berke_transmon_2022} for details),
the reason being that in a  perfectly engineered $A$-$B$-$A$-$B$ pattern, the
blocking of $A$-$B$ degeneracies would come at the price of a perfectly realized
resonance between next-next-neighbor $A$-$A$'s and $B$-$B$'s. Some degree of
frequency variation is required to prevent these next-nearest-neighbor
resonances. We are thus led to investigate a situation with weak disorder on top
of a regularly patterned background.

\begin{figure}[b]
    \centering
    \includegraphics[scale = 1]{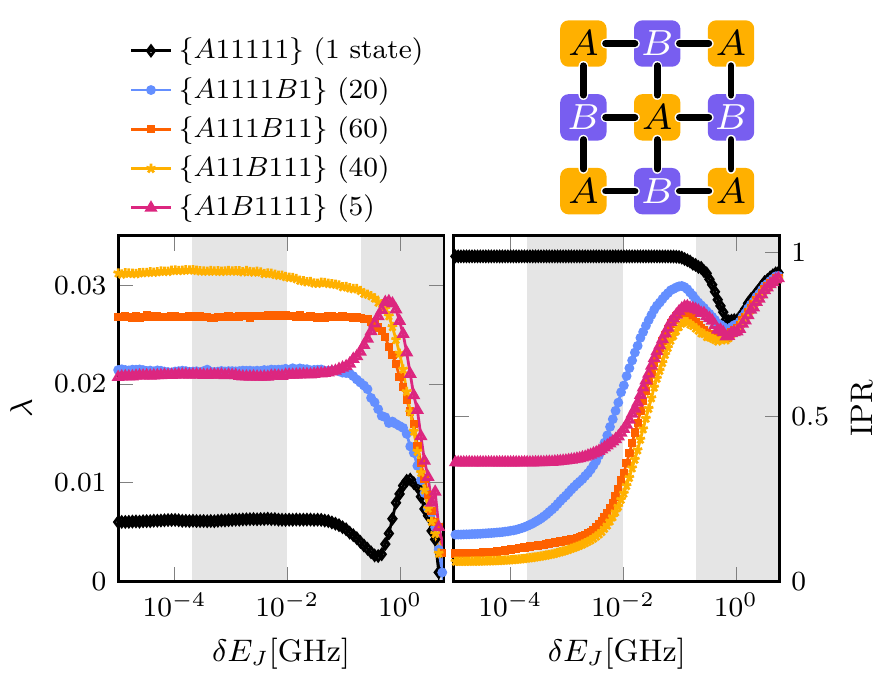}
    \caption{\textbf{State localization in precision engineered transmon arrays}
        Lyapunov exponents (left) and IPRs (right) of computational multiplets in a nine qubit $AB$ patterned transmon array averaged over 8000 realizations of disorder.($E_{J,A}
    = 12.58$~GHz, $E_{J,B} = 13.80$~GHz, $E_C=330$~MHz
    \cite{hertzberg_laser-annealing_2021} and $T=6$~MHz.)} 
	\label{fig:classicalABpattern}	 
\end{figure}

In Ref.~\cite{berke_transmon_2022}, some of us considered the  many-body physics
of a toy model of a $3 \times 3$ transmon array subject to $A$-$B$
sublattice patterning. In a nutshell, the combined effect of an order-of-magnitude
reduction of disorder and the $A$-$B$ patterning leads to a restructuring of the
Hilbert space into smaller subspaces, denoted as \emph{permutation multiplets}.
Individual multiplets harbor all Fock spaces of a definite distribution of
occupations on the two sublattices. For example, $\lbrace A11B111 \rbrace$ is
the forty-dimensional multiplet defined by all states with two $A$ sites and
three $B$ sites  in state $|1\rangle$ all others in $|0\rangle$. Multiplets are
energetically separated by energy scales defining the underlying $A$-$B$
substructure. The intra-multiplet state structure is determined by the degree of
residual disorder: from Bloch state extended (asymptotically weak disorder),
over chaotically extended (weak disorder) and intra multiplet Fock space
localized (moderate disorder), to random hybridization between multiplets
(strong disorder), see Ref.~\cite{berke_transmon_2022}.

IBM transmon engineering has managed to hit the sweet spot of intra-mutliplet
state localization, visible as a local maximum in the IPRs shown in the right
panel of \figref{fig:classicalABpattern}, where the four alternating vertical
strips represent the regimes of increasing disorder mentioned above. The  colors
represent multiplets of Hilbert space dimension between 1 and 60 defined by
the different excitation patterns indicated in the legend.  The IPR $\approx 1$  indicates a high level of
state definition for all multiplets at intermediate disorder. 

The left panel shows that the  Lyapunov exponents of this system (in its classical limit) do not
fully reveal the  quantum mechanical state structure. However, they still define
a useful diagnostic tool. Upon increasing disorder, the Lyapunov exponents
remain structureless up until the point where the optimal disorder concentration
is reached. Pragmatically, one may thus gather evidence on the preferred level
of randomness by running a Lyapunov analysis for a variety of multiplet occupation
patterns.  

\begin{figure}[t]
  \centering
   \includegraphics{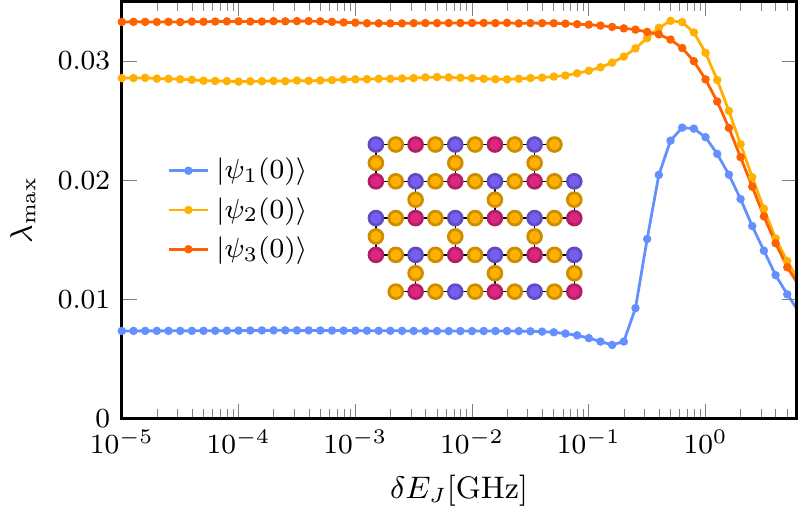}
   \caption{\textbf{Classical chaos in frequency-engineered Hummingbird chips.}
   Lyapnonov exponents of the  $C$-$A$-$C$-$B$ patterned Hummingbird chip, 
   with  $37$ ($17$) transmons in the $E_1$ $(E_0)$ state.  (Adopting IBM parameters, we set
   $E_C=330$~MHz, and  $E_J=(11.05,11.33,10.76)$ GHz for transmons on the $(A,B,C)$ sites. The coupling is  set to
   $T=6$~MHz and all results are averaged over $20{,}000$ disorder
   realizations.}
    \label{fig:classicalABC}
\end{figure}

Following the general strategy of this paper, we have pushed this analysis to
system sizes beyond the reach of quantum simulation. \figref{fig:classicalABC}
shows data for a 54-qubit $C$-$A$-$C$-$B$ patterned Hummingbird chip with 37
transmons initialized in $E_1$, and the remaining ones in $E_0$. The blue line
shows the Lyapunov exponents for the specific initial condition where the 37
excited transmons are the 37 $C$'s, i.e. a configuration corresponding to a
one-dimensional quantum multiplet. We observe  structural similarity to the
one-dimensional configuration shown in \figref{fig:classicalABpattern}, and
similarly low-lying Lyapunov exponents. For the other configurations, with random
distribution of the excited transmons, the exponents assume larger values. There is a gradual
 tendency to increased chaoticity compared to that of the smaller system, following the general trend
 observed in this paper. Finally, the more
elaborate three-frequency patterning does not appear to have a positive
effect on the stability of the system.

\section{Summary}
\label{sec:Summary}

In this paper we have proposed simulations in the classical limit $ \hbar \to 0$, but otherwise realistic system parameters, as a potent benchmarking tool for the resilience of transmon-based superconductor quantum information hardware against chaotic instability. Our analysis proceeded in a succession of four conceptual steps: 1) the demonstration of classical chaos even in small-sized arrays and energies relevant to quantum computation, 2) the construction of a correspondence showing that classical chaos evidenced by finite Lyapunov exponents implies quantum chaos evidenced by decreasing wave function participation ratios, and the extension of this correspondence to a quantitative tool, 3) the demonstration that classical simulation  is feasible for array sizes well beyond current hardware limits (and orders of magnitude beyond the reach of quantum simulations), and 4) application of this toolbox to layouts modelled after current IBM chip designs. 

The overall conclusion of this analysis is that the current engineering of
frequency-patterened transmon arrays operates in a comparatively narrow corridor
between insufficient and dangerously resonant coupling. We observe a tendency to
growing instability for larger arrays, and for growing energy
of computational states (more $|1\rangle$'s than $|0\rangle$'s.) To fully
understand the potential ramifications of chaos in this setting, it may be necessary to extend
the analysis from time independent signatures of wave functions to dynamical
protocols describing multi-qubit structures in operation. Simulating the
corresponding non-autonomous classical dynamical system will be a subject of
future research. We finally remark that additional hardware overhead, such as tunable couplers, appears to provide lasting immunization against the effects discussed in this paper. \\

\acknowledgments

We thank E. Varvelis for discussions and P. B\"onninghaus for collaboration on an initial project \cite{boenninghaus2021} 
studying incipient chaos in large-scale classical transmon systems.
We acknowledge partial support from the Deutsche Forschungsgemeinschaft (DFG)
under Germany’s Excellence Strategy Cluster of Excellence Matter and Light for
Quantum Computing (ML4Q) EXC 2004/1 390534769 (D.P.D.,S.T., and A.A.) and
within the CRC network TR 183 (project grant 277101999) as part of projects A04
and C05 (S.-D.B., C.B., S.T., and A.A.). The numerical simulations were performed on the
JUWELS cluster at the Forschungszentrum J\"ulich.


\bibliography{classicalchaos_bib}


\appendix
\section*{Appendix}

We complement our discussion of the main text with three short Appendices. 
The first one provides a compact introduction to the concept of Lyapunov exponents
in classical chaos theory, a concept routinely used in our analysis, and its
numerical computation. 
The  second Appendix gives supporting documentation of the level of fluctuations 
of Josephson energies in current-generation IBM devices.
The third Appendix provides technical background information on the algorithmic scaling 
of our numerical approach and the required compute times to simulate systems with $4{,}000$+ coupled
transmons.

\subsection{Lyapunov exponents} 
\label{app:methods}

In the main body of this paper, we quantify classical in terms of the maximal
Lyapunov exponent $\lambda$, i.e. the rate of divergence of initially
nearby trajectories. Consider the difference vector $\delta \pi = \pi - \pi'$ of
two trajectories  $\pi = \left( \vect{q}, \vect{p} \right)$ and $\pi' = \left(
\vect{q}', \vect{p}' \right)$ in the $2S$ dimensional phase space. Linearizing
the equations of motion for small $\delta \pi$ yields 
\begin{equation}
	\delta \dot{\pi} = \vect{M} \delta \pi.
\end{equation}
The matrix $\vect{M}$ contains the second derivatives of the Hamilton function
with respect to $\vect{q}$ and $\vect{p}$. With the ansatz $\delta \pi(t) =
\pi_0 \exp \left(\lambda t \right)$, one arrives at the eigenvalue equation
\begin{equation} 
	\vect{M} \pi_0 = \lambda \pi_0. \label{eq:lyapM}
\end{equation}
The eigenvalues $\lambda$ are referred to as \emph{Lyapunov exponents}. Phase space area conservation (Liouville theorem) imply that the spectrum of exponents is organized in pairs of opposite sign $\pm \lambda$, where the existence of non-zero eigenvalues is an indication of chaos. The exponent of largest modulus then determines the rate at which generic phase space separations $\delta \pi$ diverge, and therefore is the prime quantifier of chaotic instability. We refer to this maximal exponent as $\lambda$ throughout.   

\begin{figure}[h!]
    \centering
    \includegraphics{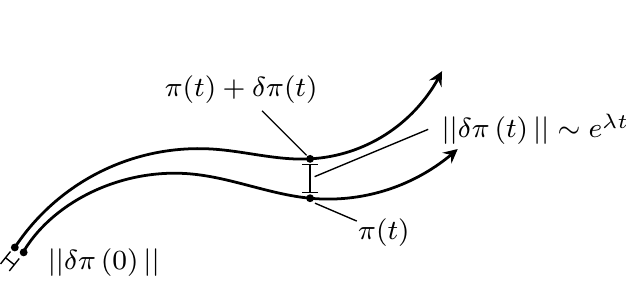}
    \caption{\textbf{Lyapunov exponents} determine the rate at which two
    trajectories  $\pi(t)$ and $\pi(t) + \delta \pi(t)$  diverge as  $|| \delta \pi(t) || \sim \exp(\lambda t)$.} 
	\label{fig:lyapunov}	 
\end{figure}

In the main text, we compute the exponents by a method proposed by
Benettin \cite{benettin1976}, where two nearby trajectories are evolved in time
and the distance vector is repeatedly rescaled at preserved
direction. The time after the divergence exceeds
a  certain phase space distance threshold $||\delta \pi||$ then determines $\lambda$, for
details see the original paper \cite{benettin1976} or Ref.~\cite{Datseris2018}.
To cross-check the results, we also compute the complete Lyapunov spectrum via an alternative method, known as `H2'~\cite{benettinLyapunovCharacteristicExponents1980a,benettinnumerics1980}.
In either case, we use the implementation provided by the software library
DynamicalSystems.jl \cite{Datseris2018} that in turn is based on
DifferentialEquations.jl \cite{rackauckas2017differentialequations}. The
equations of motion are solved using the implementation of Tsitouras 5/4
Runge-Kutta method \cite{tsitourasRungeKuttaPairs2011}. It was checked that the
results for $\lambda$ are unchanged if higher-order methods (Verner's
``Most Efficient'' 7/6 Runge-Kutta method \cite{verner_explicit_1978}), lower
error thresholds and longer evolution times (the exact $\lambda$ is
obtained as a $t \to \infty $ limit) are used.

\subsection{Disordered transmon arrays}
\label{app:disorder}

\begin{figure}[t]
    \centering
    \includegraphics[scale=.98]{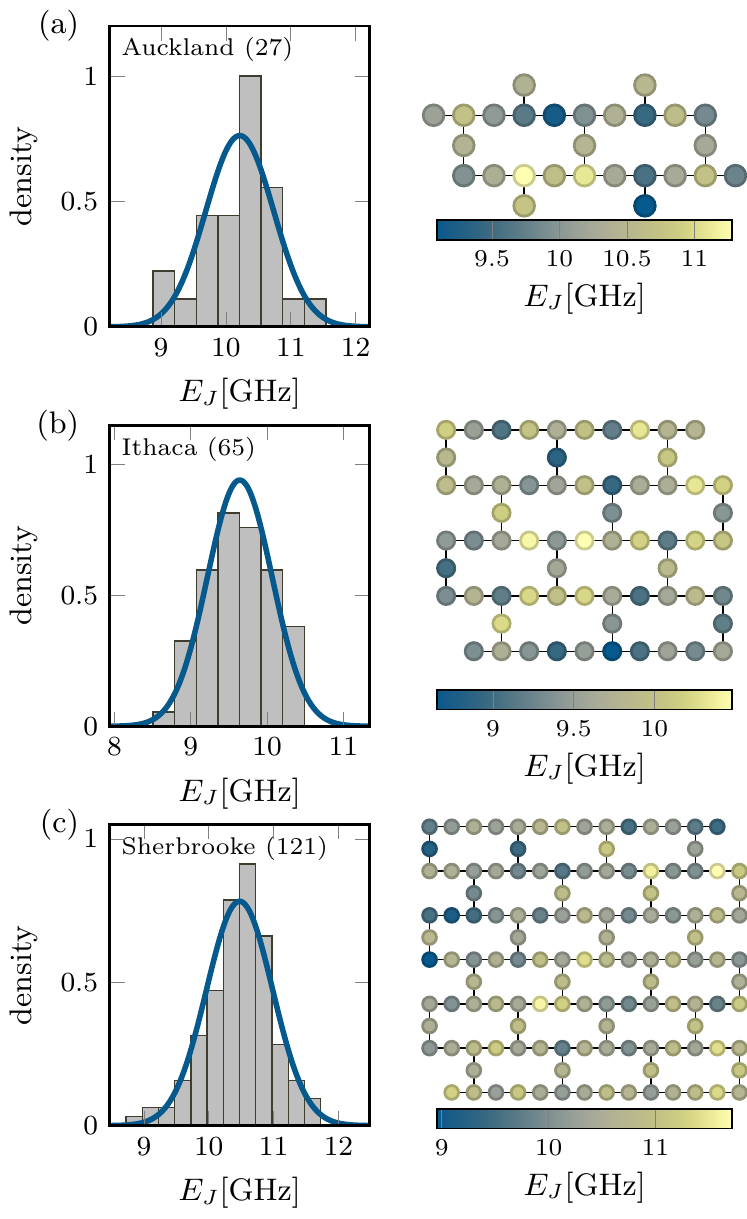}
    \caption{{\bf Examples of IBM's fixed frequency architectures.} 
    Shown is the distribution of Josephson energies for one instance of each processor generation, Falcon (27 qubits), Hummingbird (65 qubits) and Eagle (121 qubits), available in the IBM Quantum Cloud \cite{IBMQuantumCloud}. The $E_J$ spreading is consistent with Gaussian disorder (despite some post-fabrication fine-tuning). }
    \label{fig:ibmcloud}
\end{figure}

A widely used, hardware-efficient entangling gate in fixed-frequency architectures is the cross-resonance (CR) gate \cite{paraoanu_microwave-induced_2006,rigetti_fully_2010}, that switches on a $ZX$ interaction by driving one qubit with a neighboring qubit's frequency. Effective models for the CR gate \cite{magesan_effective_2020,tripathi_operation_2019} show that a small ratio of anharmonicity and qubit frequency detuning $\delta \nu_q = \nu_1 - \nu_2$, i.e., $\tfrac{E_C}{h\delta \nu_q}<1$, weakens the strength of the effective interaction and thereby slows down the gate. In state-of-the-art fixed frequency processors, one typically finds a detuning of $h \delta \nu_q \approx \tfrac{E_C}{2}$, which we take to be the definition of the disorder strength.
In the transmon regime $E_J \gg E_C$, the qubit frequencies are well approximated by $h \nu_q \approx \sqrt{8E_CE_J}-E_C$ \cite{blais_circuit_2021}, such that a scaling of the Josephson energy spread $\delta E_J$ according to $\delta E_J = \sqrt{E_J E_C/8}$, as exploited in \secref{subsec:tentransmons}, guarantees the desired frequency detuning. In addition, in the experimentally core region of $E_J \lesssim 40$~GHz, the above choice of $\delta \nu_q$ reproduces a variation of several hundred MHz in the Josephson energies, in broad agreement with typical values of the as-fabricated `natural' disorder in fixed-frequency architectures, see Fig.~\ref{fig:ibmcloud}.
For the data shown in \figref{fig:2Dvardis}, we consider larger disorders while maintaining the typical square root scaling of $\delta E_J$ with the average Josephson energy.

\subsection{Algorithmic scaling} 
\label{app:algscaling}

A difficulty in the exact diagonalization of the quantum mechanical system is
the transmon's bosonic nature leading to faster growth of the Hilbert space
dimension $n$ than the computational space of a qubit system. Even when exploiting the approximate
conservation of particle number \cite{berke_transmon_2022}, and considering
the different blocks of the Hamiltonian in \eqref{eq:fullmodel} with a fixed
total excitation number $N$ separately, the resulting matrices have dimensions
$(N+L-1)!/ ((L-1)!N !)$, where $L$ is the number of
transmons. At half-filling (the situation most commonly studied in this work and
previous studies \cite{orell_probing_2019,berke_transmon_2022}), this yields an
approximate scaling of $n \approx 2.6^L/\sqrt{L}$ for large $L$
\cite{orell_probing_2019} -- an exponentially faster growth  than the
corresponding sector with equal numbers of 0's and 1's in the computational subspace whose
dimension scales as $2^L / \sqrt{L}$ \cite{shiftinvert}. 
This naturally implies that exact diagonalization studies of
coupled transmon arrays are restricted to small systems, i.e., $L \lesssim 18$
even when using shift-invert diagonalization techniques to extract \emph{individual} eigenvectors
at high energies \cite{shiftinvert,orell_probing_2019}. For the results in
Fig.~\ref{fig:2Dvardis} we average, for each disorder realization,  the IPR over
\emph{many} eigenstates obtained by full diagonalization of the $N=5$ block of a
10-transmon chain.

\begin{figure}[b]
\includegraphics{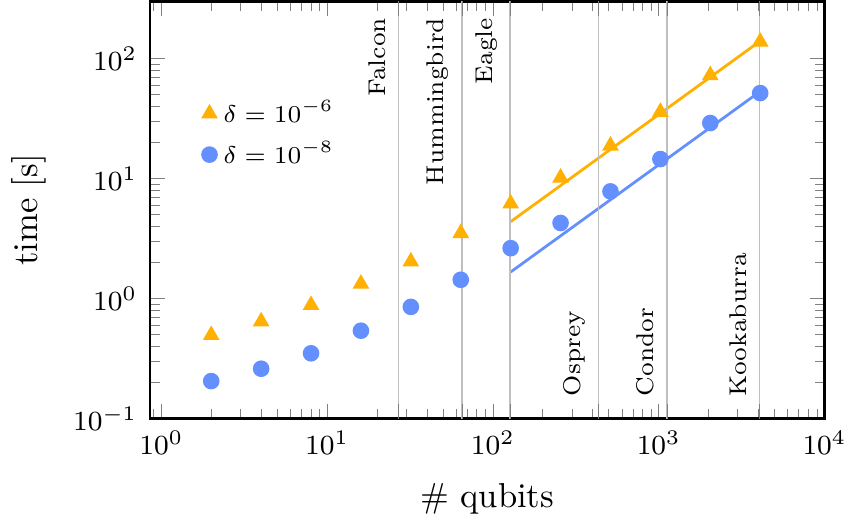}
\caption{
{\bf Compute times for the classical simulations.} Shown are measurements of the compute time per core (of an Intel(R) Core(TM) i5-8400 CPU @ 2.80GHz) of the classical simulation of different length transmon chains, for initial state $|101010...\rangle$, $E_C=250$ MHz, $E_J=10$ GHz, $\delta E_J=559$ MHz and $T=50$ Mhz, with different absolute and relative error tolerances $\delta$. The given lines are linear fits for the longest five time values of each data set, showing the asymptotic behaviour. All calculated Lyapunov exponents have $10{,}000$ underlying time steps.\\
}
\label{fig:Lyapscaling}
\end{figure}

For the classical simulation, using an explicit ODE solver \cite{rackauckas2017differentialequations} for the calculation
of the Lyapunov exponents, the effort for a single time-step propagation grows
linear in system size. Naturally, the number of steps of the differential
equation solver also enters in the computational complexity. We find
that the number of steps needed for a desired accuracy during the integration of the equations of motion does not increase with the size of the transmon array and is a roughly constant value, such that the overall computational complexity to determine the Lyapunov exponent should grow linearly in the number of transmons $L$. This is in good agreement with the asymptotic behavior for large sytem sizes shown in Fig.~\ref{fig:Lyapscaling}. Note that the numerically obtained
value of $\lambda$ approaches the exact result only in the limit $t
\to \infty$ where $t$ is the total evolution time. The computational time
increases linearly with the total evolution time $t$. Comparing the results for
$\lambda$ obtained with different $t$ ranging from $10^3$ to $5 \cdot
10^5$, we find that convergent results are typically obtained  $t \approx \mathcal{O}(10^4)$
time steps, such that, for example, analyzing one of IBM’s ‘Hummingbird’
processors containing 65 qubits takes $\approx 3.5$~s for an error tolerance
(both relative and absolute) of $10^{-8}$ in the integration of the differential
equations.

\end{document}